### Title
- **Title**: Ants, robots, humans: A self-organizing, complex systems modeling approach
- **Short title**: A self-organizing, complex systems modeling approach

### Authors
M. Jaraiz[1]*

### Affiliations
[1] Department of Electronics, University of Valladolid, Valladolid 47011, Spain.
*Corresponding author. Email: mjaraiz@ele.uva.es

### Abstract
Most of the grand challenges of humanity today involve complex agent-based systems, such as epidemiology, economics or ecology. However, remains as a pending task the challenge of identifying the general principles underlying their self-organizing capabilities. This article presents a novel modeling approach, capable to self-deploy both the system structure and the activities for goal-driven agents that can take appropriate actions to achieve their goals. Humans, robots, and animals are all endowed with this type of behavior. Self-organization is shown to emerge from the decisions of a common rational activity algorithm, based on the information of a system-specific goals dependency network. The unique self-deployment feature of this approach, that can also be applied to non-goal-driven agents, can boost considerably the range and depth of application of agent-based modeling.

**One Sentence Summary:** An agent-based approach capable to self-deploy the system structure and simulate the movie-like activity of each individual.

## MAIN TEXT

### Introduction

"What makes James Bond an agent? He has a clear goal, he is autonomous in his decisions about achieving the goal, and he adapts these decisions to his rapidly changing situation". This introductory sentence, from a Grimm *et al. (1)* review on agent-based modeling (ABM) of complex systems, provides a vivid definition for an agent. According to the review we are surrounded by such autonomous, adaptive agents: plants, cells of the immune system, citizens, stock market investors, businesses, etc. One of the most important challenges confronting modern science is to understand and predict complex systems such as ecology *(1)*, epidemiology *(2, 3)*, or economics *(4)*. The review proposes a framework that may lead toward unifying algorithmic theories of the relation between adaptive behavior and system complexity. However, the identification of general principles underlying the organization of ABSs has been hampered by the lack of an explicit strategy for coping with the two main challenges of bottom-up modeling: complexity and uncertainty. A recent editorial article from a special issue on meeting grand challenges in ABSs *(5)* reiterates the challenge of identifying general principles underlying the systems' internal organization. Similarly, in the realm of economics, previous studies *(6)* have pointed out that there is an old but still unresolved concern of economists, such as the Nobel prize Hayek *(7)*, about what are the self-organizing capabilities of decentralized market economies. Regarding the 2008 global crisis, the president of the European Central Bank, in his opening speech at the 2010 ECB Conference, clearly described the problem: "Allow me now to turn to the broader issue of lessons from the crisis for macroeconomics and finance. When the crisis came, the serious limitations of existing economic and financial models immediately became apparent. (…) Macro models failed to predict the crisis and seemed incapable of explaining what was happening to the economy in a convincing manner. As a policy-maker during the crisis, I found the available models of limited help. In fact, I would go further: in the face of the crisis, we felt abandoned by conventional tools. (…) Agent-based modelling dispenses with the optimization assumption and allows for more complex interactions between agents. Such approaches are worthy of our attention." *(8)* Perhaps it was too difficult to predict the economic crisis, but the worst thing was being incapable of explaining what was happening to the economy during during all those years.

As an answer to the quest about the general principles underlying self-organization in complex agent-based systems, this article presents an activity deployment simulation scheme (ADS), Fig. 1, that can self-deploy both the system structure and the agents activity of any ABS whose agents satisfy the above given definition.

**Results**
As a demonstration, we have implemented a proof-of-concept program (see supplementary materials for a detailed description) that automatically deploys all the necessary production and marketing activities of each individual in a colony of settlers (humans *(9)* or robots *(10-12)*) that arrive to an uninhabited island and build up a small State from scratch, over several decades. Now we sketch the grounding ideas of the simulation approach. As a complementary approach to another already used *(1)*, based on observed patterns that result from the activity of the system and trying to infer the model structure, we have explored the reverse path: define the possible actions of the constituent agents, set their goals, add as a complement the network of interdependencies of those goals and assess the outcome of the activity. To facilitate the explanation with an example we take Robinson Crusoe as the agent, to deal initially with tangible goods such as houses and windows. Let us suppose that the agent wants to have a house and thinks "For a house I need 4 windows, 3 doors, 40 m of wall and 5 months of work" and then "For a window I need 0.5 $m^2$ of glass, 1.5 m of wood frame and 0.5 days' work" etc. So, he first goes to the forest to get the wood, then collects sand and coal to make glass, then assembles the windows, etc. If there are several agents in the island, Crusoe can ask another neighbor, called Friday, for coal and then they agree that Friday, that also wants to have a house, will produce the coal and sand while Crusoe will produce the wood. This naive line of chained production and marketing activities turns out to be a common pattern of activity for adaptive, goal-driven agents, whether human, robots or honeybees. From it emerges the concept of a universal activity deployment simulator as sketched in Fig. 1.

The same ADS code can coordinate the execution of subcommands by a robot to accomplish several intertwined tasks (Fig. 1, see supplementary materials for details). Complex agent-based systems are, thus, shown to emerge from the decisions of a rational activity algorithm (RAA), Fig. 2, common to all goal-driven agents, based on information from a system-specific goals dependency network (GDN). This combination conforms the activity deployment simulator (ADS), that besides answering Hayek's question, is capable to build the system architecture and deploy the coordinated activity of agents as disparate as bees *(13)*, or a robot like a human operator in a laboratory *(10)*, a robots colony *(15)*, or the response mechanisms to infection from pathogen agents *(16-18)*. The activity deployment simulator, ADS, operates the rational activity algorithm, RAA, on the goals dependency network, GDN, to achieve the agent's goals. This dependency network arises from the different abstract activity patterns (i.e., actions) that each agent can perform. Another valuable feature of this modeling framework, besides self-deployment, is that the level of detail of the GDN's modular architecture can be adapted and increased locally in the network as needed, to better describe the mechanisms relevant to the problem at hand, while other more global network nodes can represent less critical mechanisms. The example provided by the proof-of-concept program is a minimal implementation of the model. For example, behavioral economics criteria, *(19)* drawn from psychology, could be applied to drive agents' choices under different circumstances. It is on those choices (seen by the agent as 'needs', however illogical) where the RAA algorithm would operate.

The activity deployment simulator opens the way to a new type of simulation, that of the integral human activity. For instance, and as illustrated with the settlers colony example, from the monthly or daily activity of individuals and industries it would be possible to estimate the amount and geographical distribution of generated $CO_2$, *(20)* pollution, their effects on climate that in turn affect production and activities, spread of an epidemics through direct interaction between specific individuals and its influence on their activity, to mention just a few. As a more specific example, in the current turmoil of the COVID-19 pandemic, where health and economics dominate the agenda, governments around the world are relying on computer simulations to help guide decisions aimed at minimizing its global impact *(3, 21, 22)*. As discussed by Epstein *(2)* on previous similar cases, agent-based models played an important part in designing containment strategies for smallpox in 2001. They helped to shape the avian flu policy, and later, in the 2009 swine flu pandemic, played a central role in mapping the disease's possible spread and designing policies for its mitigation.

However, current ABMs operate on synthetic populations of individuals *(23)* that must be preconfigured, and this is a particularly sensitive issue because the outcome can depend critically on a properly initialized, self-consistent population. Due to the lack of a general framework, the model structure is often chosen ad hoc. The ABM approach presented here can introduce a new perspective: the self-consistent and coherent simulation of the spread of a pandemic throughout an ADS-deployed network of interactions between individuals in their daily labor and social activities, including naturally all sorts of population heterogeneities *(24)*. For instance, regarding protection against future zoonosis outbreaks, a clear link has been identified between deforestation and virus emergence because tropical forest edges are a major launchpad for novel human viruses acquired from contact with wildlife *(25, 26)*. The ADS approach can provide the most realistic and detailed ABM simulations, including land use and the population interacting with it, as the simple example of settlers suggests, if developed on a specific geographic map. Thanks to the continuous update of the system structure changes that result from the individuals' activities in the ADS simulations, other pivotal questions could also be analyzed, such as, will this decision convey permanent damage to the economic fabric or only transitory stress? What are the consequences of two alternative measures? *(27)*

In addition, as noted above, the modular GDN network architecture makes it ideally suited to tackle complex systems through collaboration between teams specialized in different fields -another desirable feature *(28)*- such as industry, finance, epidemiology, or environmental sciences. Although it has a different impact in each country, a global pandemic can best be suppressed by global collaborations and this agent-based simulation approach can be used to leverage the knowledge of specialized teams across the world.

Regarding the simulation size, already in 2009, the Global-Scale Agent Model (GSAM) included 6.5 billion agents and executed an entire US run in around ten minutes *(2)*. A stochastic agent-based model of the SARS-CoV-2 epidemic in France was run recently on 500,000 individuals *(21)*. The Republic of Mauritius is a small island nation in the Indian Ocean, with a population of 1.3 million people and an area of 1,865 km2. An integrated analysis of climate change, land-use, energy and water strategies *(29)* was used by its government to define the relevant policies. The program presented here simulates the 10-year monthly activity of 100.000 individuals in 16 minutes on a single thread of a laptop computer, and this proof-of-concept version can still be highly optimized.

**Discussion**

The self-deployment and self-organization features of the ADS framework can also be exploited in non-goal-driven systems whose complexity arises from the architectural complexity of their interaction network. For instance, for the immune system cells the goal can be just random movement or, upon a given interaction, it can switch to follow the gradient of a certain chemical. It is from the richness of interactions and responses from where the crowd wisdom of the immune system is going to emerge. As an example of application, discussed in the supplementary materials, the same ADS code can coordinate the execution of subcommands by a robot to accomplish several intertwined tasks *(11, 12)*. In this case, the program creates a population of virtual individuals (homunculus) and assigns to each one the responsibility of the execution of one of the subcommands. Each homunculus negotiates the timely execution of its subcommand through free market interactions with the rest of them. In the example of a robot assembling a chair, to attach the 'right' side of the chair (Fig. 1) to the leftframe assemblage, the corresponding homunculus has to find in the market another one (a seller) that has already assembled several other parts to form the leftframe intermediate structure. In both examples, Crusoe's goods and robot's actions, goal 2 requires to first carry out goals 3 and 4 and that is reflected in the abstract (but specific for each application) goals dependency network of Fig. 1. However, the reasoning (decision making) is the same in both cases, the deployment algorithm sketched in Fig. 2. These simplistic ideas can drive the self-organization of complex adaptive systems, like the settlers example shown in Fig. 3 and discussed in detail in the supplementary materials. This simplicity was already anticipated by Grimm *et al. (1)*: "Theories of complex systems may never be reducible to simple analytical equations, but are more likely to be sets of conceptually simple mechanisms (e.g., Darwinian natural selection) that produce different dynamics and outcomes in different contexts". In our case, the simple mechanisms correspond to the universal RAA and the different contexts correspond to the different GDNs, that are specific for each problem and dictated by the nature of the goals.

Finally, the unique self-deployment capabilities of the ADS approach, as displayed in the settlers example, can expand the range of application of agent-based modeling. For instance, it could easily generate complex settings and activities to provide background live stages for the vast market of video games and movies. Other fields can be expected to benefit from this highly detailed, dependency-driven simulations of complex real-world systems.

**Materials and Methods**

As discussed above, the ADS modeling approach (Fig. 1) can be applied to different fields. We have chosen economics to implement a detailed example of its application (a settlers colony). Given the length of the description of the implementation details, - although basic and accessible to readers from any field- they have been included in the supplementary materials. The simulation scheme (and the program itself) is very simple, the complexity of the outcome arises from the richness of the GDN dependency network. As a brief comment about its implementation, the program iterates for a given number of steps (timesteps or counter) through the agents and, at each step, each agent tries to acquire its current needs/goals from its interacting neighbors, following the RAA algorithm, Fig. 2.

**Competing interests:** Authors declare no competing interests.

**Figures and Tables**

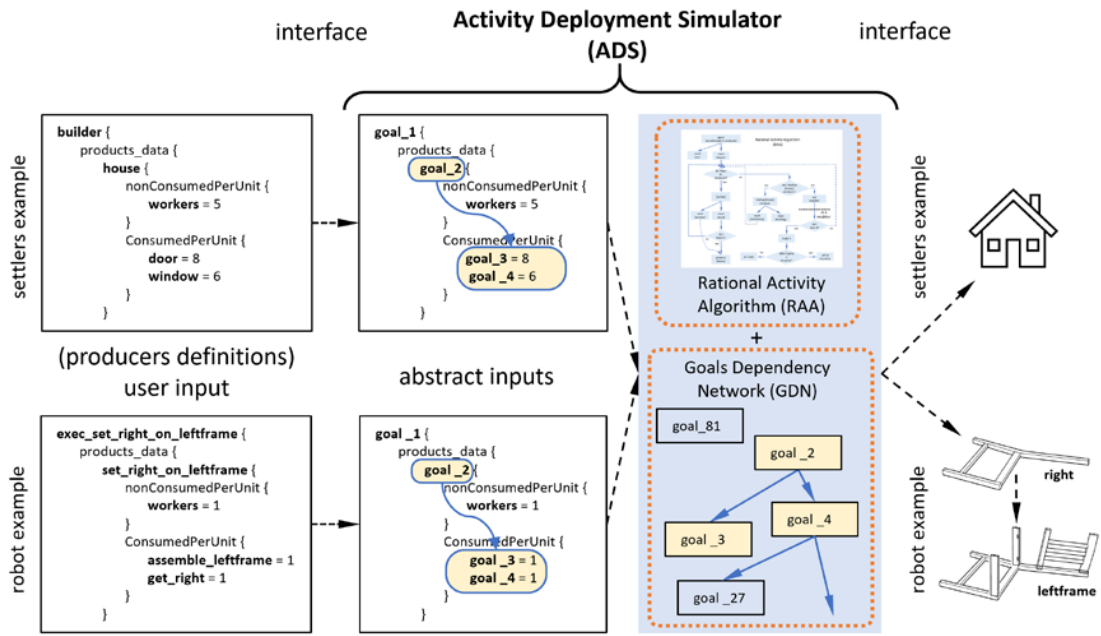

**Fig. 1. Activity deployment simulator (ADS).** Agents take appropriate actions to achieve their goals. The dependence of each goal on its means translates into an abstract Goals Dependency Network (GDN) that defines the specific system architecture. The activity of rational agents arise from simple and *universal* rules: the *driving forces* are the wants of each agent that trigger the *deployment* of production chains (defined by the system's specific GDN) under the decisions of the Rational Activity Algorithm (RAA). The *self-organizing* capabilities of complex systems and societies stem from the RAA + GDN architecture. For each application, the ADS core engine also needs an application-specific interface, as shown. The GDN level of detail can be adjusted locally as needed: in the settlers example, 'window' can be an end node or be replaced by 'assemble-window' and the corresponding sub-goals, like those shown in the robot example.

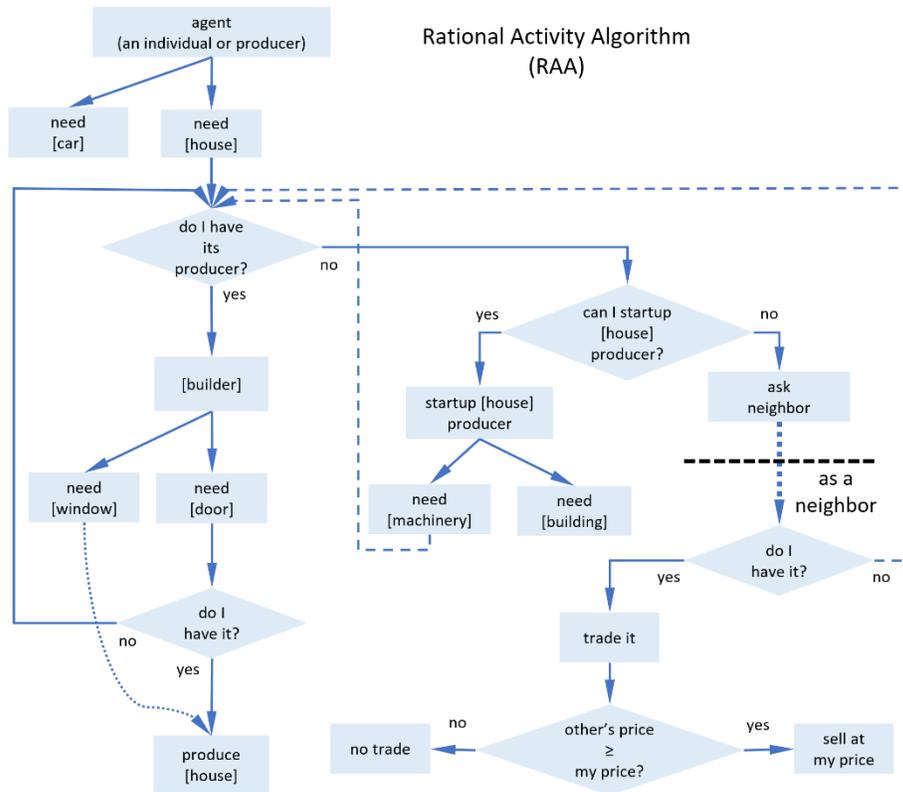

**Fig. 2. Rational activity algorithm (RAA).** Rational agents have to adapt their decisions to changing situations. In the case of collaborative agents, if the agent cannot produce a needed good, it assigns a subjective *price* to that good, relative to its other needs, and tries to get it from other agents. After a successful trade, the seller (buyer) increases (decreases) its price to improve its own benefit at the next trade. Failed trades induce the reverse price changes. Collaborations without money (robots, bees) omit the price checking step. This simple mechanism of acting as a collaborative neighbor can give rise to collective, supra-agents (such as a bee swarm or an ant colony) that can perform far more complex tasks than their constituent, isolated individuals.

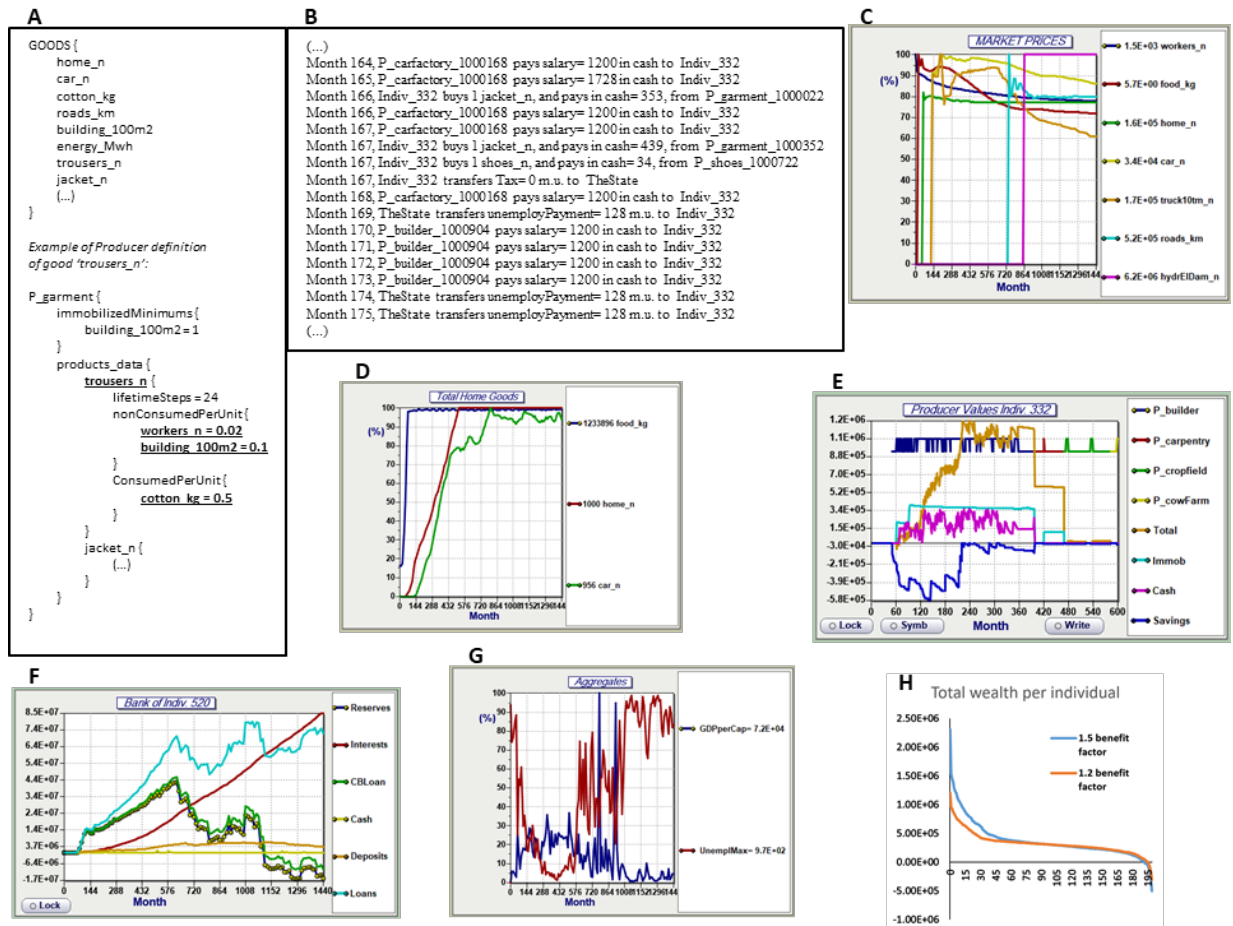

**Fig. 3. Self-deployment of a complex production and marketing system in a simulated colony of settlers.** From just the producers' physical properties and individuals' wants (**A**), a wealth of concerted and productive activity emerges with realistic timescales and prices as the output of an ADS simulator. (**B**) Example of output log. (**C**) Time evolution of some market prices. (**D**) It takes several decades to build all the houses and cars. (**E**) Example of output relative to the producers of one individual. (**F**) Example of output from a Bank. (**G**) Visualization of some macroeconomic aggregates (Gross Domestic Product per capita and Unemployment). (**H**) The wealth of individuals exhibits a Pareto distribution, and an increase in the sales benefit factor increases only the benefits of the wealthiest individuals, in agreement with theoretical models *(14)* that show that inequality is driven primarily by chance, rather than by differential individuals' abilities (here all individuals are identical).

**Supplementary Materials**

Other supplementary materials for this article include the following:

> *Deployers* program and input file examples.

**Supporting Information: Application examples**

Example 1: Economics

To simplify the explanation of the modeling approach we assume, as an example, the following scenario: a governor, herein referred to as the State, is sent with thousands of settler households to an uninhabited island, arriving initially with only some hand tools. Since they come from a civilized country, they are supposed to bring already all the necessary know-how to make and build whatever they need. The governor brings enough money, whose use and appreciation is already familiar to the settlers, and establishes a Central Bank. The aim of each settler is to have food, clothing, a house, and a car. The State also wants some buildings, cars, roads, and a hydroelectric dam built.

The program implements four types of agents: individuals, Producers (firms), Banks, and the State. To run the simulation (1000 settlers, 120 years) rename a copy of the DeployersInput-Settlers-120years.txt file to DeployersInput.txt. This file should be in the same directory as Deployers.exe and DeployersInterface.g. The simulation takes about 100 s on a 4GHz computer.

*Production*

The problem is how to plan and organize all this development. To give an overall description of how the program (*Deployers*, Fig. S1) works we begin with a simplified example of input file (Fig. S2). The input file, generated by the program user, must include a list with the Goods (possible goals) names, that can be arbitrary although, to improve readability, it may be helpful to append the units (_kg, _km...). The task required from the program user is the definition of Producers (the businesses, not the owners), listed also in the input file. Producers are goods themselves, but for simplicity they are not traded in this version. They play a key role but, to simulate the activity, it is not necessary to know their internal operation, only what we can call their *activity pattern*. For example, a corn crop has a different activity pattern than building a house or making a car because, unlike the house or the car, the crop requires a fixed number of months, regardless of the number of workers. For this proof-of-concept version we have implemented a basic activity pattern template with several variables (number of workers, duration...) that allow the simulation of some patterns. Expanding the template in future versions would make it a more 'universal' ADS. As an example of producer definition, the garments producer of Fig. S2 lists two products: trousers and jackets. From the trousers definition we see that 0.5 kg of cotton is consumed per trouser. The *nonConsumedPerUnit* workers_n = 0.02 entry implies that one worker can produce up to 50 trousers per month. Also, 100 $m^2$ surface building is needed for that production rate. The *immobilizedMinimums* entry means that, for production rates needing less than 200 $m^2$ surface, a business building is not necessary. This reflects the early stages of development with, for example, individuals working with just a few workers to build the first houses whereas, later on, a large building company will need to get a large enough building or several trucks before starting high-rise building constructions.

The *Control_Variables* allow the user to change some parameters to try this first version of the program. Some hints are included as comments in the input file itself. Figure S3 shows a few lines of an output file. The program, using a one-month timestep, iterates *nYears* over a randomized list of the *nIndividuals* to let them perform their *MonthlyActivity* (Fig. S4):

Products with a lifetime (see trousers in Fig. S2) get worn out and need to be replaced after a certain time (amortization), with a random time uncertainty margin (*ProductsDecay()*).

Every month each individual gets a random list of interacting individuals and producers (businesses) for marketing operations. The list of desired goods (food, clothing, a house, and a car) is also defined in the user input file. According to their current needs, using that list as the target, they proceed to make a list of goods to buy and then go to *BuyGoods()* from their interacting neighbors.

Then the individual turns to care about his businesses (producers). Those deemed to be inefficient are closed down. In this version, the yardstick is a maximum time of inactivity, *MaxUnusedTimestps*, although real accounting criteria could be applied because individuals get data collected during the simulation (Fig. S3) that is analogous to that of real life and could maintain double-entry account bookkeeping. The goods (buildings, machinery...) from a dismantled producer can then be re-used for starting a new producer. Otherwise they are put on sale.

The way the program implements entrepreneurial initiatives of the individuals (*TryToStartupNewProducer()*) starts by 'suggesting' a producer type at random (carpentry, bakery...). But the idea is discarded immediately if there is an input good that is not yet available in the market. Otherwise, the individual still waits to get some product demand in the market before starting to invest and produce.

The last monthly operation of individuals is to launch their producers' *MonthlyActivity* task (Fig. S5). In this case, the producer (an agent itself) asks its interacting 'neighbors' (individuals and producers) the quantities of its products that they want to buy, makes a list of the input goods needed for production, tries to buy them (this includes hiring employees) and runs production with the available input goods. Afterwards, it loops again through the neighbors list to offer and sell the available product quantities. Then, if possible, returns loans to its Bank, transfers benefits -if any- to its owner and pays taxes to the State at the end of year.

In this initial version of the program, the quantities of goods and money are integer numbers. This means that the minimum non-zero quantity is 1 unit, and this should be considered when deciding the goods units. For this reason, for example, we use land_10m2 to have fractions of one hectare.

In short, the way the simulation proceeds is the following: one individual makes the list of goods to buy, for example trousers, and tries to purchase them from his neighbors. If neighbor1 happens to have the hint or 'suggestion' of starting a garment factory, neighbor1 goes to the Bank to get a loan, asks another neighbor for cotton, this in turn asks another for land, and so on. Thus, the first products to appear in the market will be those available from the natural resources that only need work, like wheat, cotton, wood from the forest, and coal and metal from mines. The next manufactured goods to appear sequentially can be garments, carts, windows, houses, machinery, cars and trucks, highways, hydroelectric dams and so on.

Each producer, after a production stage, computes and updates its production price of that good as the cost of input goods plus salaries, divided by the number of units produced, and multiplied by a benefit factor (*PriceBenefitPer100*). Market prices are recorded for output every month as the average of the transaction values during that month.

The value of the monetary unit (m. u.) is not defined directly by the user. It arises as a result of the individuals' salary demand: in the simulation they are set to start asking for a random salary value between 500 and 5000 m. u. With the producer definitions given in the example input file, the market salary value settles at around 1200 m. u. Figure S6 shows the time evolution of the salary (*workers_n*) and several goods. Legend's values are peak prices. The market salary starts at around 1500 m. u. and settles at 1200. The vertical lines correspond to the time of first appearance of each good in the market. Note that food is fully available almost from the beginning, then houses start to be built and cars later on because they need large machinery to be manufactured. It takes several decades to build all the houses and cars (Fig. S7). The fluctuations in the number of cars arise from their lifetime (amortization) input parameter: they get worn out and need to be replaced from time to time. Thus, the correct timing in the production sequence can arise from the interdependence of goods, provided that the definitions of their producers correspond to their real-life properties.

Macroeconomic aggregates can easily be analyzed with this detailed microeconomics approach. Unemployment (Fig. S8) is minimum during the years where houses and cars are still under production (rising curves in Fig. S7) but there is already enough infrastructure to start working on the large roads and the single dam constructions for the State. In this last phase, the unemployment is high. The annual Gross Domestic Product (GDP) per capita is also shown. The timing of the goods produced for the State is shown in Fig. S9.

*Market*

Individuals and producers trade their goods following simple criteria. For example, a producer should not sell its products at a price below the fabrication cost. However, this limit can be ignored, if necessary, when dismantling a producer. A trade is carried out if the buyer price is higher than or equal to the seller's. The price agreed upon is the seller's. Another rule is that both learn from each marketing trial, whether failed or successful: they readjust their price in the opposite direction with a factor (*PriceAdaptDefault*) defined in the input file. If *bFixedPrices* is set to 1, prices remain fixed at their first value.

*The State*

The State, represented by the Governor, owns the Central Bank and, initially, all the money. From the beginning (see GOODS_I_WISH in the input file) the State tries to purchase from the public market a list of items (buildings, cars, roads, and a large dam) as representative of public expenditures. On the other hand, the State collects a direct tax (personal income tax) and an indirect tax (value added tax, VAT). Finally, as automatic stabilizers, the State pays an insurance of unemployment and uses a progressive rate to charge the personal income tax (see Fig. S10). This modeling framework allows to easily implement complex economics policies.

*Banks*

The money source is the Central Bank (CB), that is owned by the State. The total amount is defined in the input file with the *MonetaryBasePerCapita* parameter, together with CB liability and asset rates. Individuals are allowed to get loans and make deposits in the CB, subjected to those rates, only until private commercial Banks begin to appear.

The Central Bank authorizes individuals to establish private commercial Banks if they have a minimum threshold balance (*BanksInitCapital*) in their CB account and requires those Banks to maintain at least a given *ReserveRatio* of their clients' deposits. In this version, the liability and asset rates that commercial Banks apply to their clients are increased by a *BanksRateFactor* over the CB values. With the ADS modeling approach, it is easy to implement and simulate any of the operations available in real financial markets (bonds, shares, Stock exchange...). Figure S11 shows the time evolution of some Central Bank values. The monetary base in this example is 1.0E8 m. u. (Reserves) but Loans are above that quantity because of the 1% *ReserveRatio* that allows for a much higher monetary supply. Figure S12 displays how the monetary base (coins plus bank notes) is distributed throughout the course of the simulation. Payments can be done in cash or by Bank transfers (see printout of an individual's activity, Fig. S3).

Commercial Banks use the CB to draw or deposit money (deposits: negative CBLoan values in Fig. S13). For simplicity, interests obtained by the Bank are being used to provide further loans. Generally, at least a fraction would be transferred to the owner or distributed to the shareholders. A fuller implementation of the program should include a financial market to allow inter-bank transactions, as well as competition between them to gain investors by offering a wide financial asset portfolio. Thus, the program could be used to test different strategies employed by both Banks and investors.

*Individuals*

Figure S14 displays data of individual 332 (type 332 in the Indiv text box and hit return) as a representative case of the self-organized sequence of activities that arise from the agent's behavior, in response to the changing conditions of the environment. The *Producer Values* plots indicate, with changing colors at the top and listed in the legend, the sequence and activity of the individual's producers. To simplify visualization, individuals are not allowed to own more than one producer at a time. Producers that failed to start production are not plotted, and we see that to be the case for the first 5 years of this 50-year simulation. This individual run four producers sequentially. The most productive (*Total* wealth curve) was the building company and the large loan (negative *Savings*) was cancelled after month = 200, but that business was closed when there was no more demand for houses. Many of these strategies are commonly implemented in current agent-based models. What is novel in this approach is the story-like deployment of activities, triggered by the individuals' wants and needs to accomplish their goals, throughout the simulated time span. The *Individual Values* plots, that also include at the top a representation of working months, show that this individual worked during the first 5 years for someone else. The *Total* wealth curve is the sum of cash and all personal goods, including the house and car. Going back to the Producer Values plot, after closing down the building business, the immobilized machinery appears in his *Home Goods* plots and is sold soon after as can be seen in the DeployOutput file. Other businesses are then tried: carpentry, crop field and cow farm. A partial listing of the activity of Indiv_332 can be seen in Fig. S3. It is remarkable that, although all individuals in the simulations presented here are identical, their fate follow quite different tracks and end up with a few very wealthy, a few very poor and a broad middle class (Fig. S15), in agreement with the theoretical result *(14)* that inequality is driven primarily by chance, rather than by differential abilities of the individuals. Interestingly, the two simulations of Fig. 3H in the main text suggest that an increase in the benefit factor (sell price / production cost per unit) increases the wealth of only the wealthiest individuals. Innate or acquired differences between individuals could be incorporated in the model by assigning 'genetic' parameters to them and could also be optimized with a genetic algorithm. And, to simulate the rise of a civilization, invents and discoveries can be simulated by just lowering the probability for an individual to get a hint to start producing a new type of good (wheel, cart, car) so that it takes many years for that good to start being produced.

This 120-year, 1000 individuals simulation takes less than 2 minutes and uses less than 1 GB memory on a single thread of a 4 GHz laptop computer. This includes a large overhead common to all simulation sizes. For example, the simulation of 10 years with a population of 100.000 individuals takes only 16 minutes and 3 GB, and the program can now be redesigned to be more efficient.

Example 2: A robots colony

A robots-only colony can be simulated by avoiding failed trade attempts that arise due to lack of money from the buyer or disagreements on price, so that the quantity traded is directly the minimum of consumer demand and producer supply.

To run a simulation of the settlement with robots instead of individuals, rename a copy of the DeployersInput-Robots-120years.txt file to DeployersInput.txt. This file should be in the same directory as Deployers.exe and DeployersInterface.g. Compared to the settlers case, the only input parameters changed (left-shifted for easy identification) are those needed to get price agreement in all trades and to remove taxes and other money related issues. Alternatively, a similar effect can be obtained by simply setting *bRobotsColony* to 1 in the settlers input file. The initial unemployment falls very quickly because there are not price disagreements and all the work gets done in a much shorter time than with the settlers. The simulation takes about 95 s on a 4GHz computer.

Example 3: A single robot

As another example, this approach can also be applied at a different level, to command the actions of a robot. This level of applicability is based on the equivalence between producers (carpentry) that generate goods (chair, table) and tasks (*AssembleChair*) that generate actions (*PickPin, LocateHole, InsertPin*). Note that producers are also goods themselves and, likewise, whole tasks can be seen as single actions in an upper level of abstraction. Within a robot, that can perform many different tasks (*AssembleChair, AssembleTable*...) using low level actions (*PickPin*...), the ADS approach can be used to generate a valid sequence of actions in response to the assigned tasks and to the changing conditions of its environment. For example, a robot that has been asked to assemble a chair *(11)* and set up a table for dinner *(12)*, using the necessary pieces as they become available in a random order, Fig. S16, is analogous to a garment factory that is asked to produce a number of goods (jackets, shirts...): in both cases they can proceed performing their chores in parallel as the needed inputs become available. According to ref. 10 the sequence of steps was hard coded through a considerable engineering effort. To run a simulation of the control of a robot, rename a copy of the DeployersInput-SingleRobot.txt file to DeployersInput.txt. This file should be in the same directory as Deployers.exe and DeployersInterface.g. The simulation takes less than a second. The command sequence is listed at the end of the _DeployersOutput.txt file. In this case, the input file sets the variable *bSingleRobot* to 1, to get a specific output format. But to verify that the program is operating as in the previous examples, the same robot commands can be executed with the DeployersInput-Robots-Assemble.txt file that is a slightly adapted version of the DeployersInput-Robots-120years.txt file, essentially in the GOODS_I_WISH section. The program assigns the execution of each command (producer) to a different agent (individual). Individual_0 is asked to assemble the chair and set the table.

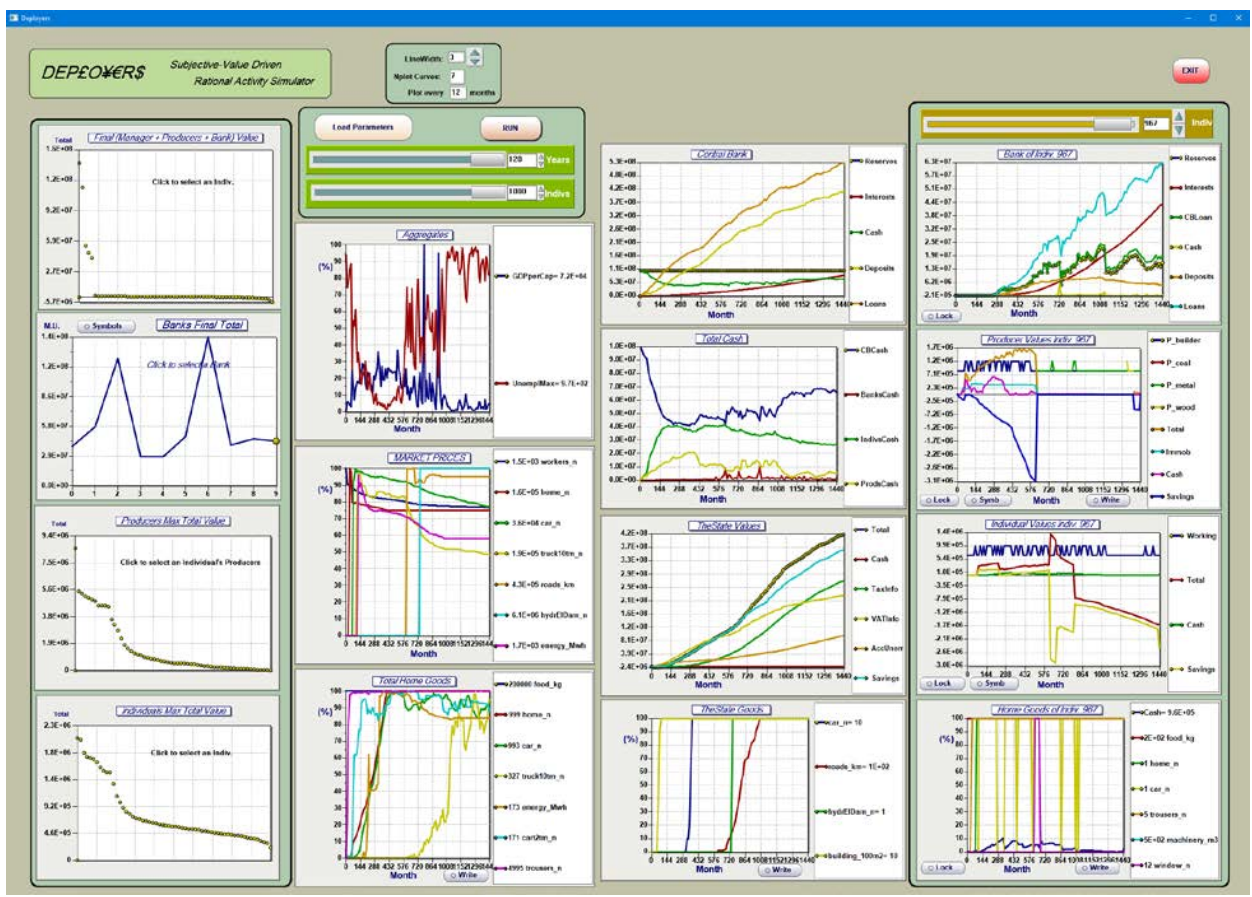

**Fig. S1.**

**Overall view of the *Deployers* graphical user interface.** In this case, the underlying simulation engine, based on the ADS approach, is used in the field of Economics.

```
GOODS {
workers_n food_kg window_n home_n car_n
door_n jacket_n cotton_kg wheat_kg roads_km
hydrElDam_n
}
PRODUCERS {
    P_garment {
        immobilizedMinimums {
            building_100m2 = 2
        }
        products_data {
            trousers_n {
                lifetimeSteps = 24
                nonConsumedPerUnit {
                    workers_n = 0.02
                    building_100m2 = 0.02
                }
                ConsumedPerUnit {
                    cotton_kg = 0.5
                }
            }
            jacket_n {
                ...
            }
        }
    }

    P_builder {
        ...
    }
    ...
}

CONTROL_VARIABLES {
    120  nYears
    2000  nIndividuals
    40  nMaxInteractingNeighbors
    120  PriceBenefitPer100
    120  PriceAdaptDefaultPer100
    100000  MonetaryBasePerCapita
    1  LiabilityRatePer1000
    3  AssetRatePer1000
    200000  BanksInitCapital
    2  BanksRateFactor
    1  ReserveRatioPer100
    15  VATPer100
    20  TaxPer100
    10  PayUnemplPer100
}
```

**Fig. S2.**

**Simplified example of input file.** Taken from a settlers colony example.

```
(…)
Month 164, P_carfactory_1000168  pays salary= 1200 in cash to  Indiv_332
Month 165, P_carfactory_1000168  pays salary= 1728 in cash to  Indiv_332
Month 166, Indiv_332  buys 1 jacket_n, and pays in cash= 353, from  P_garment_1000022
Month 166, P_carfactory_1000168  pays salary= 1200 in cash to  Indiv_332
Month 167, P_carfactory_1000168  pays salary= 1200 in cash to  Indiv_332
Month 167, Indiv_332  buys 1 jacket_n, and pays in cash= 439, from  P_garment_1000352
Month 167, Indiv_332  buys 1 shoes_n, and pays in cash= 34, from  P_shoes_1000722
Month 167, Indiv_332  transfers Tax= 130 m.u. to  TheState
Month 168, P_carfactory_1000168  pays salary= 1200 in cash to  Indiv_332
Month 169, TheState  transfers unemployPayment= 128 m.u. to  Indiv_332
Month 170, P_builder_1000904  pays salary= 1200 in cash to  Indiv_332
Month 171, P_builder_1000904  pays salary= 1200 in cash to  Indiv_332
Month 172, P_builder_1000904  pays salary= 1200 in cash to  Indiv_332
Month 173, P_builder_1000904  pays salary= 1200 in cash to  Indiv_332
Month 174, TheState  transfers unemployPayment= 128 m.u. to  Indiv_332
Month 175, TheState  transfers unemployPayment= 128 m.u. to  Indiv_332
Month 176, TheState  transfers unemployPayment= 128 m.u. to  Indiv_332
Month 177, TheState  transfers unemployPayment= 128 m.u. to  Indiv_332
(…)
```

**Fig. S3.**

**Example of some printout data from Individual 332.** The time step is set to one month but an agent can carry out many activities on the same month. Payments can be done in cash or through a Bank transfer. The State collects taxes and VAT, and pays unemployment.

```
CIndividual::MonthlyActivity()
{
        // At home: my private goods and activities:

        MonthlyFeed();
        ProductsDecay();// Products with lifetime decay by use
        MakeListOfInteractingIndivsAndProducers();
        MakeListOfGoodsToBuy();
        BuyGoods();

        // At work: run my businesses (Producers):

        ClosedownInactiveProducers();
        TryToStartupNewProducer();
        for myProducers:
                Producer.MonthlyActivity();
}
```

**Fig. S4.**

**Simplified MonthlyActivity tasks of Individuals.** In the *ProductsDecay* function, goods with a *lifetimeSteps* input parameter get worn out and have to be replaced.

```
CProducer::MonthlyActivity()
{
        MakeListOfInteractingIndivsAndProducers();
        MakeListOfDemandsForMyProducts();
        MakeListOfGoodsToBuy();// input goods, based on products demand
        BuyGoods();
        RunProducer();
        SellMyProductsToInteractingNeighbors();
        ReturnLoansToMyBank();
        TransferBenefitToManager();
        if (_pGeneration->CurrentMonth == 11)
                PayAnnualTax();
}
```

**Fig. S5.**

**Simplified MonthlyActivity tasks of producers.** In *MakeListOfGoodsToBuy* the Producer makes an estimate of production size based on recent market demand, to make the list of input goods that have to be purchased for production.

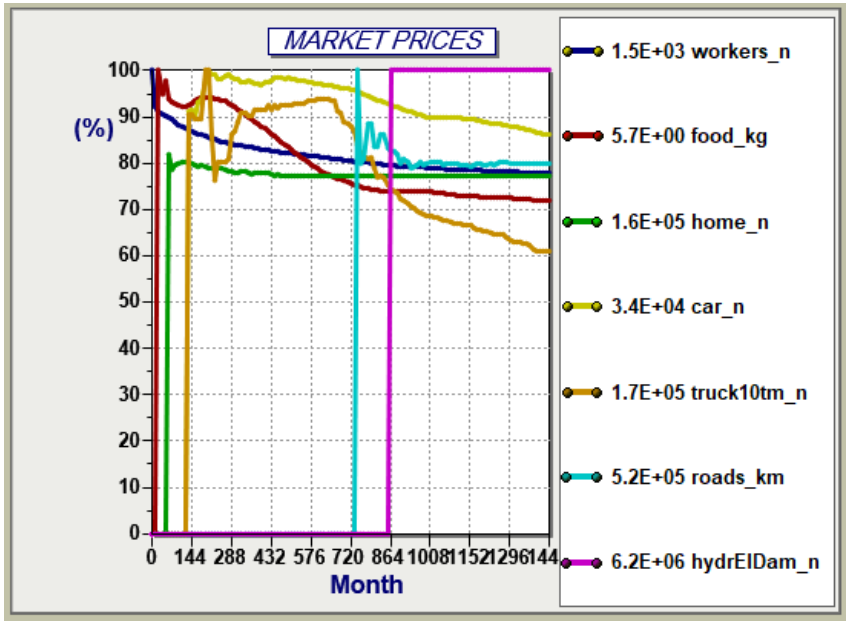

**Fig. S6.**

**Market prices showing the time sequence of first appearance of some goods in the market.**

Legend values are peak prices. Workers_n corresponds to the salary.

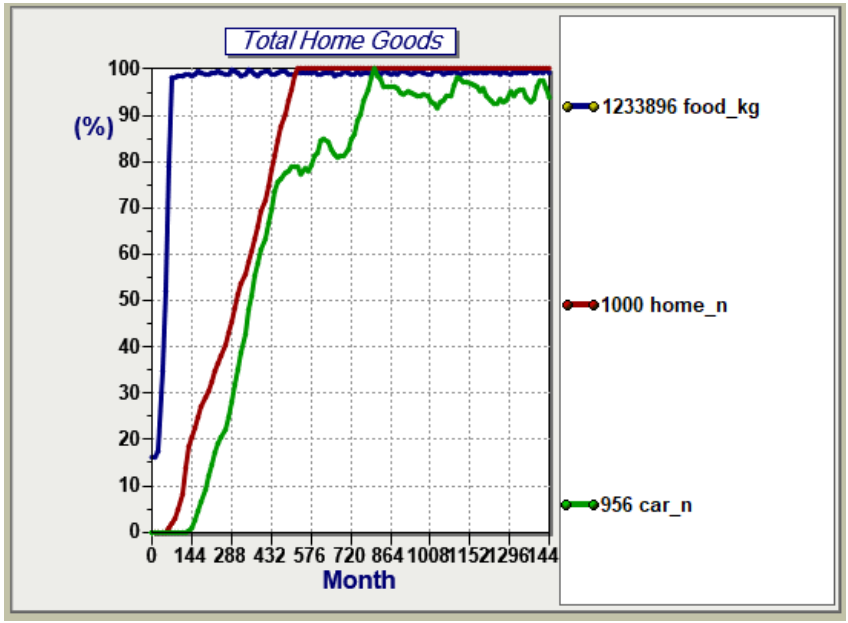

**Fig. S7.**

**Production units of some individuals' goods vs time.** It takes several decades to build the houses and cars because all industrial infrastructure has to be set up for the first time. Brief declines in the number of cars are due to their lifetimeSteps parameter.

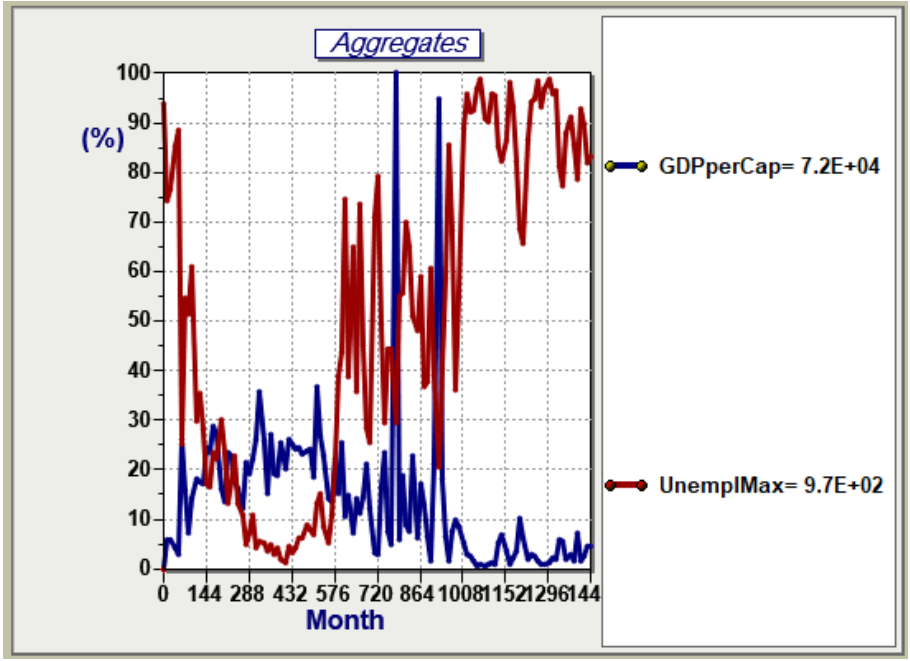

**Fig. S8.**

**Unemployment and Gross Domestic Product (GDP) per capita.** It takes time to achieve full employment at the beginning until market prices settle down and many producers are in operation.

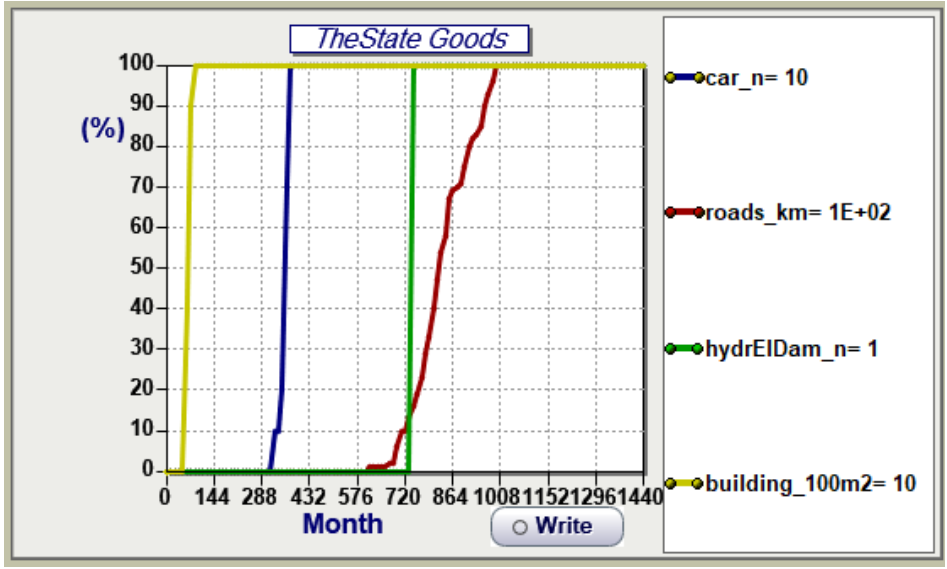

**Fig. S9.**

**Production units, for the State, of buildings, cars, roads, and a dam vs time.** Large constructions, like highways and dams, can only appear after all the required industrial infrastructure has been set up.

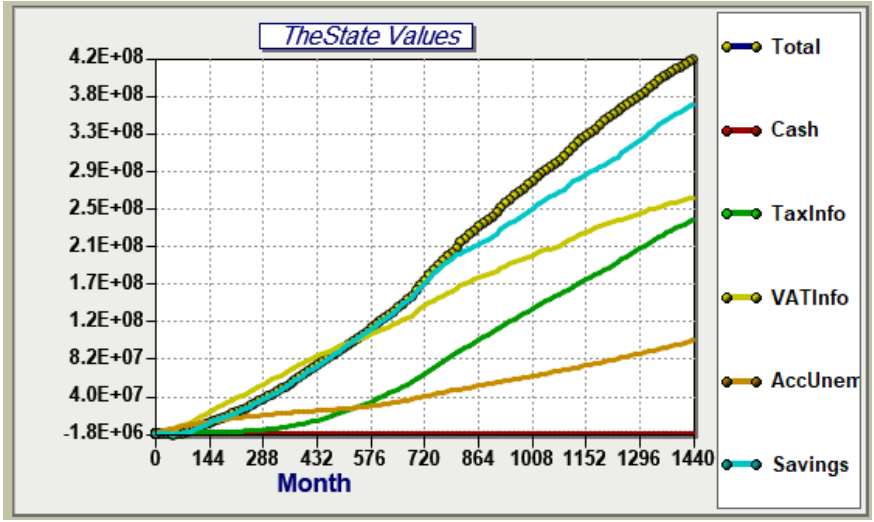

**Fig. S10.**

**Some State variables vs time.** Total = Cash + Savings (in the CB) = TaxInfo + VATInfo – AccUnempl.

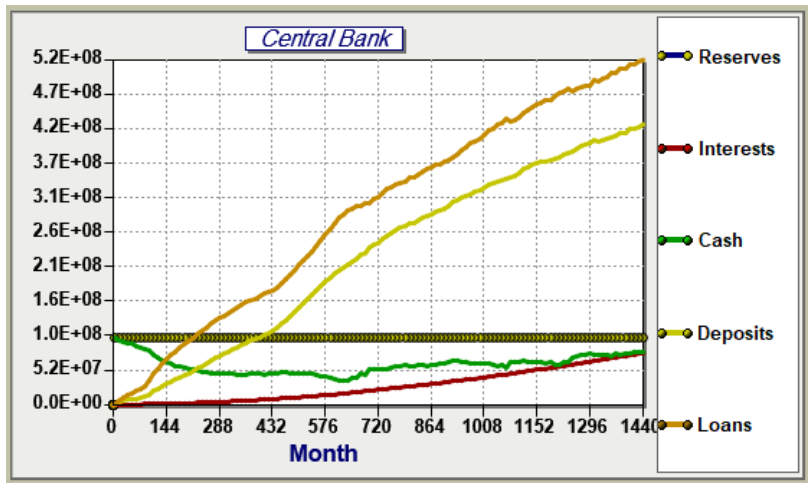

**Fig. S11.**

**Time evolution of some Central Bank data.** The monetary base in this example is 1.0E8 m. u. (Reserves) but Loans are above that quantity because of the 1% *ReserveRatio* that allows for a much higher monetary supply.

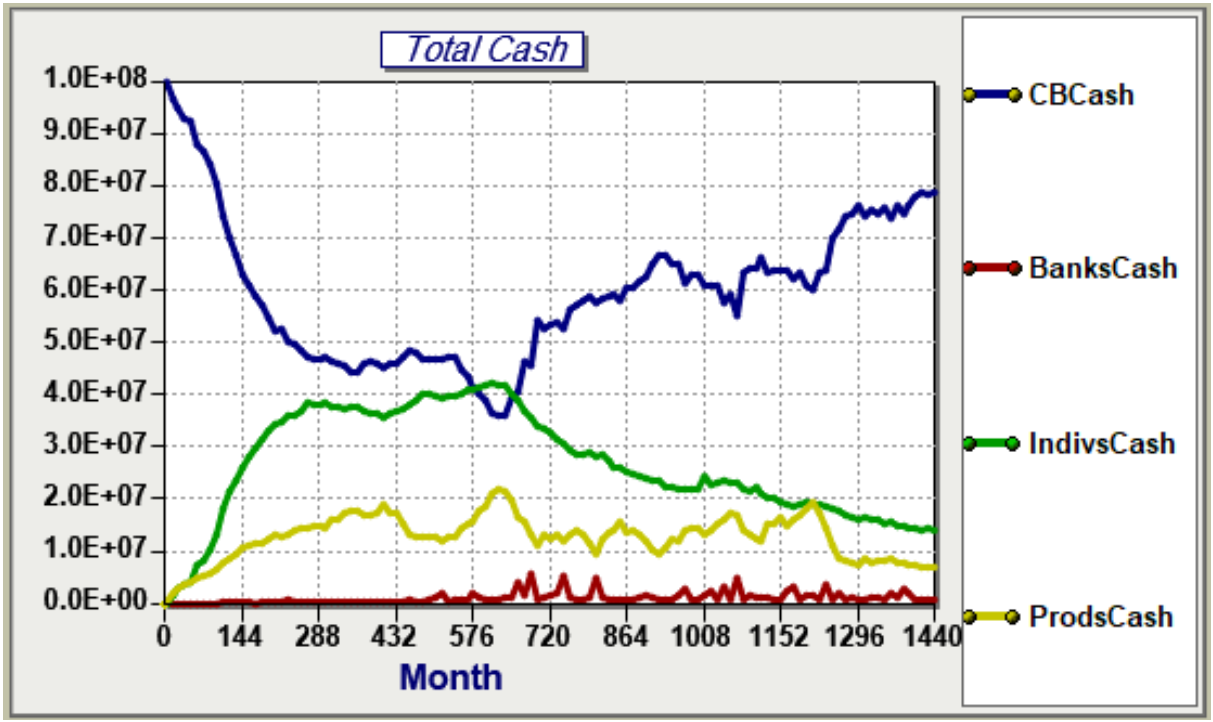

**Fig. S12.**

**Distribution of the monetary base.** Monetary base = coins + Bank notes.

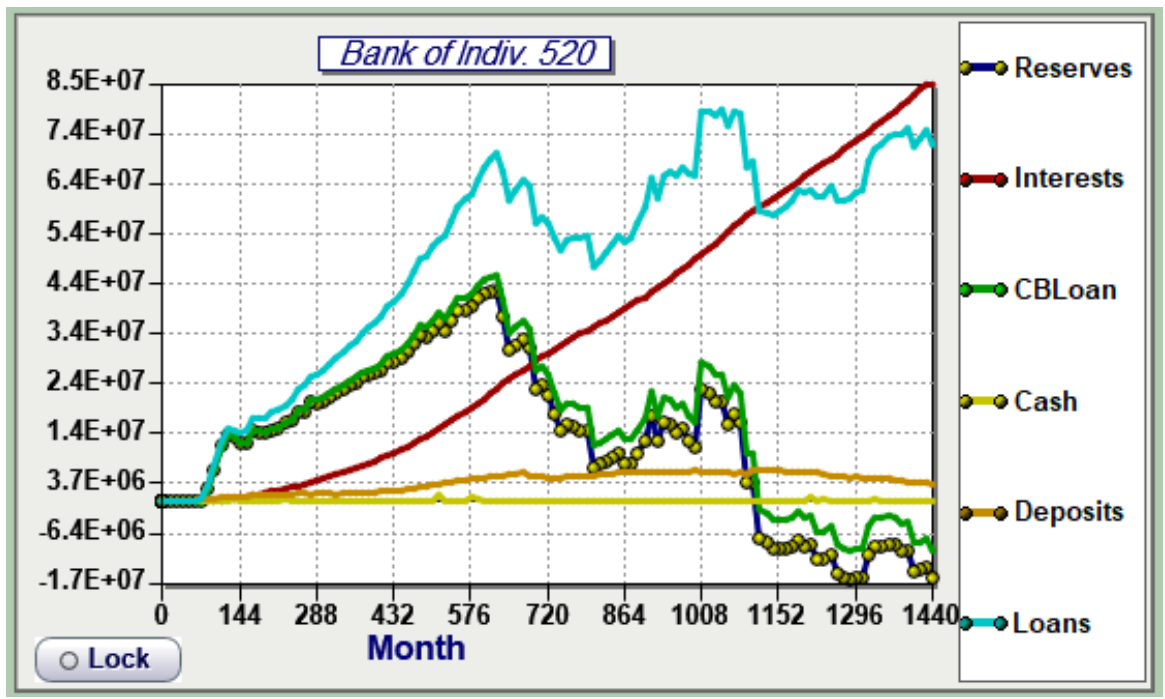

**Fig. S13.**

**Data from a private commercial Bank.** Loans to clients are the main source of profitability (Interests). Negative CBLoan values correspond to positive deposits of this Bank in the CB.

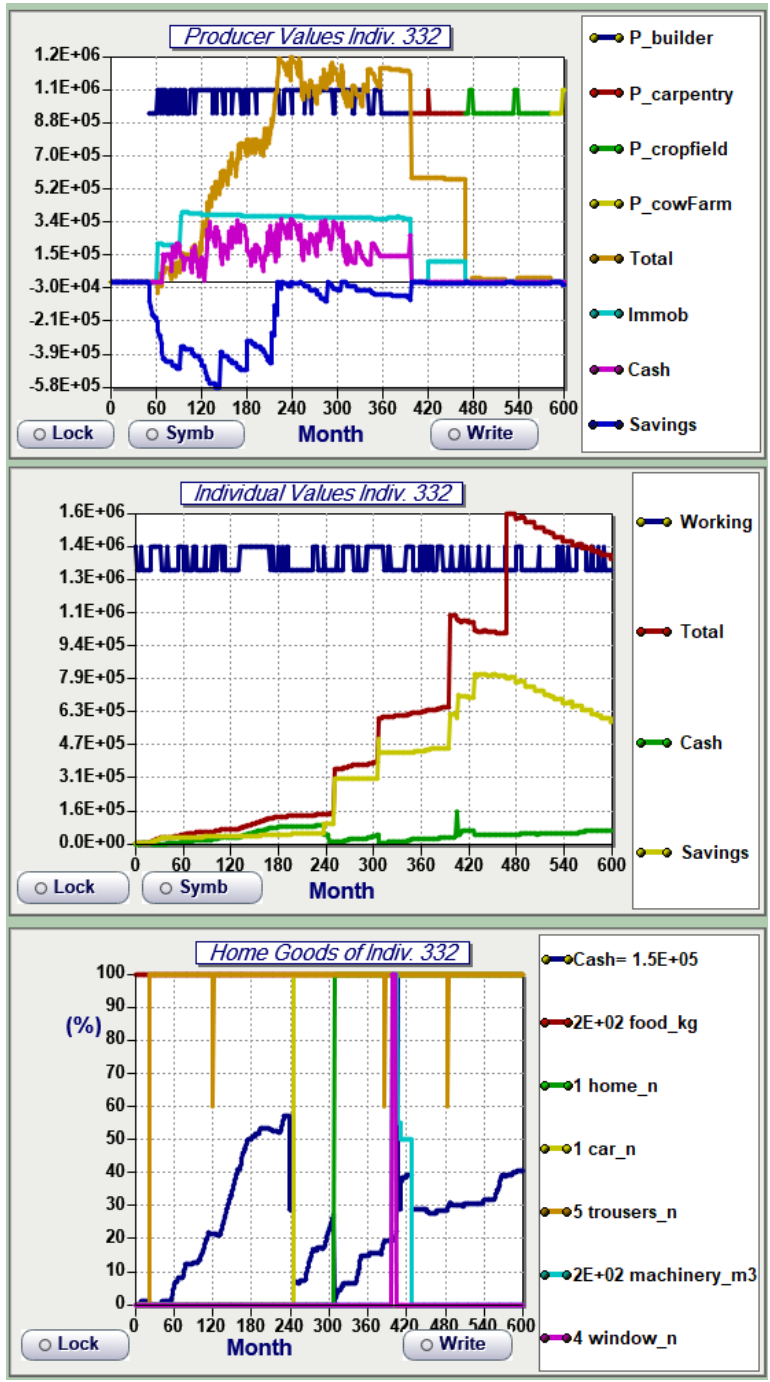

**Fig. S14.**

**Data from individual 332.** The Producer Values plots indicate, with changing colors at the top and listed in the legend, the sequence and activity of the individual's producers. Total = Immob + Cash + Savings + StoredGoods (not shown). The sharp Total fall near Month 480 corresponds to transfers of StoredGoods to the owner (see Total of Individual Values plots).

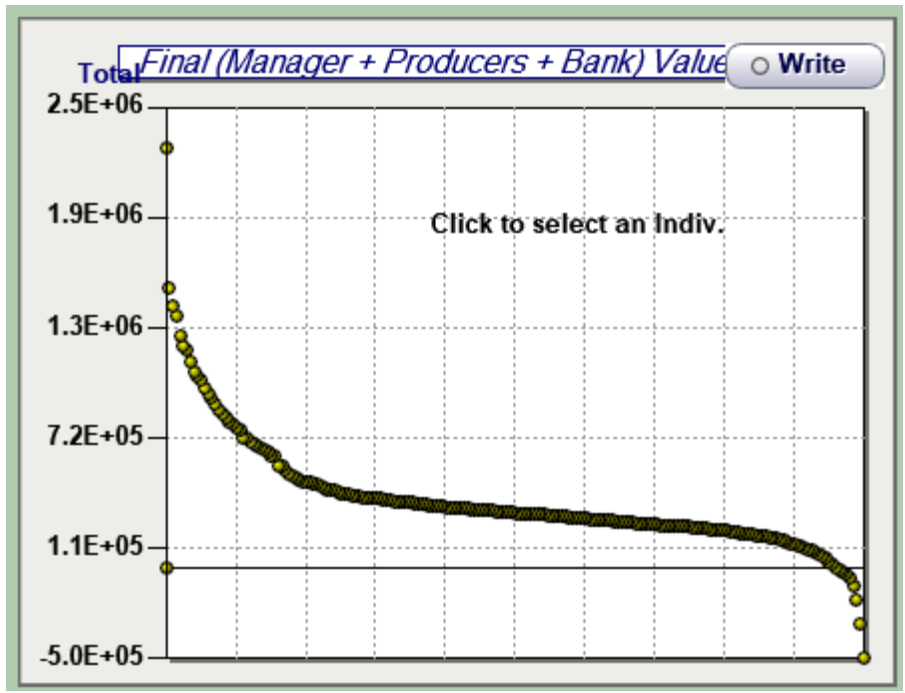

**Fig. S15.**

**Plot of individuals' wealth.** Sampled from the full population, ordered by decreasing total final value of individual plus Producer plus Bank, if any. To get this smooth curve set *MaxNBanks* to 0.

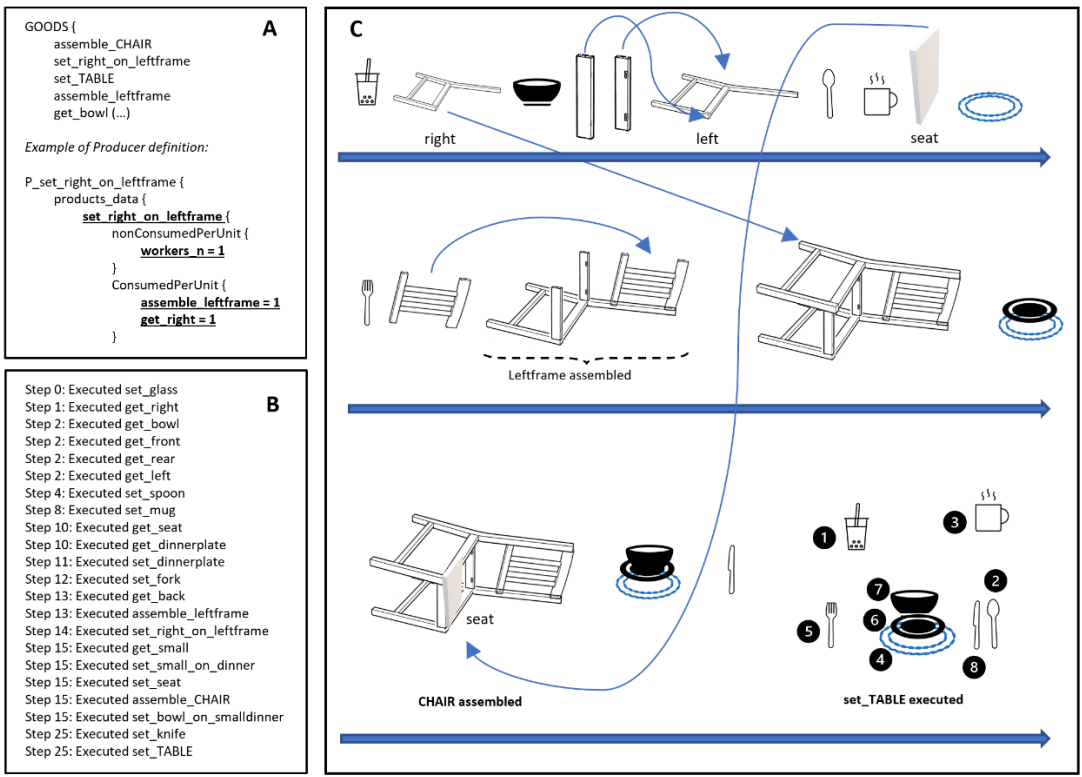

**Fig. S16.**

**Robot control.** For an agent, sequencing of (intangible) tasks is equivalent to production of (tangible) goods. The same program that deploys a complex production system can orchestrate the timely execution of subtasks in a robot that has been commissioned to perform many different tasks simultaneously, as the means become available is an arbitrary order. Items arrival sequence left to right and top to bottom. The 'right' chair side is not used until the left frame has been assembled in the second row. The table setting needs to lay the dinner plate, small plate, and bowl in that order. (A) Excerpts from the input file. (B) Output: robot's actions sequence generated by the ADS simulator. (C) Visualization of the actions corresponding to the output sequence.